\documentclass[12pt,thmsa]{article}
\usepackage{sw20aip}

\bigskip
\input{tcilatex}
\begin{document}

\title{Numerical Investigation of a Bifurcation Problem with Free Boundaries
Arising from the Physics of Josephson Junctions}
\author{M. D. Todorov \\
%EndAName
{\small \textit{Institute of Applied Mathematics and Computer Science,}}\\
{\small \textit{Technical University of Sofia, Bulgaria}}\\
{\small \textit{e-mail: mtod@vmei.acad.bg}}\\
\\
T. L. Boyadjiev\\
{\small \textit{Faculty of Mathematics and Computer Science,}}\\
{\small \textit{Sofia University 'St. Kliment Ohridski', Bulgaria}}\\
{\small \textit{e-mail: todorlb@fmi-uni.sofia.bg}}}
\date{}
\maketitle

\begin{abstract}
A direct method for calculating the minimal length of ``one-dimensional''
Josephson junctions is proposed, in which the specific distribution of the
magnetic flux retains its stability. Since the length of the junctions is a
variable quantity, the corresponding nonlinear spectral problem as a problem
with free boundaries is interpreted.

The obtained results give us warranty to consider as ``long'', every
Josephson junction in which there exists at least one nontrivial stable
distribution of the magnetic flux for fixed values of all other parameters.
\end{abstract}

\section{Posing the Problem}

It is known that the stationary distributions of the magnetic flux $\varphi
(x)$ in ``long'' (one-dimensional) Josephson junctions (JJ) are solutions of
the nonlinear boundary value problem (BVP)

\begin{eqnarray}
-\varphi _{xx}+j_{D}(x)\sin \varphi +\gamma\!\!\!& = &\!\!\!0\>,\quad x\in
(-R,R)\>,  \label{eq:1} \\
\varphi _{x}(\pm R)\!\!\!&=&\!\!\!h_{B}\>,  \label{eq:2}
\end{eqnarray}

\noindent where $h_{B}$ is the external magnetic field alongside the axis $y$
on the junction plane (see Fig.$1a$).

We note that the kind of boundary conditions (\ref{eq:2}), either in
presence or absence of current $\gamma $ in the right side of eq.(\ref{eq:1}%
), are determined by the geometry of the junction. Here we consider simple
junctions with overlap geometry \cite{barone}, in which the current $\gamma $
can be approximately considered as a constant. The generalization of our
results in cases of any other geometry, for example in-line geometry, does
not require big efforts.

We suppose that the given continuous function $0\le j_{D}(x)\le 1$ describes
the variations of the Josephson current amplitude, caused by the possible
local inhomogeneities of the dielectric layer thickness. When $%
j_{D}(x)\equiv 1$ the junction is homogeneous. Otherwise when the junction
is inhomogeneous the function $j_{D}(x)$ is usually modelled by Dirac $%
\delta $ - function \cite{barone,galpfil} or its continuous approximations,
for example hyperbolic functions \cite{boyad85}, splines \cite{alexboyad}
etc. At the present work as is in \cite{alexboyad}, we use an isosceles
trapezium with base $\mu $ (see Fig.$1b$) as more suitable in physical sense
model of inhomogeneity.

Every solution of the equation (\ref{eq:1}) is simultaneously a stationary
solution of the perturbed Sin-Gordon equation (SGE)

\begin{equation}
\varphi _{tt}+\alpha \varphi _{t}-\varphi _{xx}+j_{D}(x)\sin \varphi +\gamma
=0\>,  \label{eq:3}
\end{equation}

\noindent where $\alpha $ is coefficient of a resistance. When $\alpha >0$
the second term in the above equation is dissipative and hence arbitrary
distribution $\varphi (x,t)$ of the magnetic flux as result of an energy
loss can be ``attracted'' by some steady distribution $\varphi (x)$.

In order to study the stability of some concrete solution $\varphi (x)$ of
the BVP (\ref{eq:1}),(\ref{eq:2}) we consider the following Sturm-Liouville
problem (SLP)

\begin{equation}
-\psi _{xx}+q(x)\;\psi =\lambda \,\psi \>,\quad x\in (-R,R)\>,  \label{eq:4}
\end{equation}
where $q(x)=j_{D}(x)\cos \varphi (x)$ is a potential, originated by the
solution $\varphi (x)$, and boundary conditions of Neumann's type

\begin{equation}
\psi _{x}(\pm R)=0\>.  \label{eq:5}
\end{equation}

It is well known that the SLP of such kind contains a counting set of
different eigenvalues 
\[
\lambda _{min}\equiv \lambda _{0}<\lambda _{1}<\lambda _{2}<\;...\;\lambda
_{n}<\;...\>, 
\]
and every one corresponds to a unique eigenfunction $\psi _{n}(x)\>,$ $%
\>n=0,1,2,\ldots $ , determinated by the norm condition

\begin{equation}
<\psi ,\psi >\,\equiv \int_{-R}^{R}\psi ^{2}(x)\;d\,x=1\>.  \label{eq:6}
\end{equation}

If the minimal eigenvalue $\lambda _{min}>0\>$, the respective solution $%
\varphi (x)$ of BVP (\ref{eq:1}),(\ref{eq:2}) is stable with respect to
small time-space perturbations. If the minimal eigenvalue $\lambda
_{min}<0\> $, this solution is unstable. The eigenvalue $\lambda _{min}=0$
is a bifurcation point, in which the stable solutions of eq.(\ref{eq:1}) go
to unstable ones and vice versa (for details see \cite{galpfil}).

Apart from the space coordinate $x$, the virtual solutions of the nonlinear
BVP (\ref{eq:1}),(\ref{eq:2}) depend also on the physical parameters $h_{B}$%
, $\gamma $ and ``technological'' ones $\mu $ and $R$ , i.e., $\varphi
=\varphi (x,p)$, where we simply substitute $p\equiv \{h_{B},\gamma ,\mu
,R\} $. The varying of every of those parameters causes a variation of the
distribution $\varphi (x,p)$ and therefore subsequent variations of the
potential $q(x,p)$, the eigenvalues $\lambda (p)$ and the respective
eigenfunctions $\psi (x,p)$. Thus we can conclude that every solution of BVP
(\ref{eq:1}),(\ref{eq:2}) has an area where it remains stable with regard to
the variations of the parameters $p$. The equation

\begin{equation}
\lambda _{min}(p)=0  \label{eq:7}
\end{equation}

\noindent determinates in the parametric space a hypersurface which points
appear to be bifurcation points corresponding to the solution under
consideration (\ref{eq:1}). The intersections of the bifurcation
hypersurface (\ref{eq:7}), when there are fixed pairs of the parameters $p$,
we call bifurcation curves and respective values of the parameters -
bifurcation (critical) parameters. The most interesting from the physical
viewpoint seem to be bifurcation curves ``external current - magnetic flux'' 
$\lambda _{min}(h_{B},\gamma )=0,$ when the geometrical parameters $\mu $
and $R$ are fixed. That curves could be relatively easy obtained
experimentally \cite{vystavkin}.

From the technological point of view, however, it is worth investigating the
bifurcation curves as functions with respect to at least one of geometrical
parameters

\[
\lambda _{min}(h_{B},R)=0\>,\text{ }\>\lambda _{min}(\gamma ,R)=0\>,\>\text{
or }\lambda _{min}(\mu ,R)=0\>. 
\]

Such kind of problems can be connected with the optimization of sizes of
devices, containing Josephson junctions.

Formally the numerical modelling of the bifurcation curves as function of
some concrete parameter $p_{0}\in p$ (at the paper \cite{boyad86} it is
chosen $p_{0}=h_{B}$) can be schemed as follows. We find some solution of
the BVP (\ref{eq:1}), (\ref{eq:2}). After that we check his stability with
respect to small perturbations solving the corresponding SLP (\ref{eq:4}) - (%
\ref{eq:6}). If $\lambda _{min}>0$ we set a new value $p_{0}:=p_{0}+\Delta
p, $ where $\Delta p$ is given increment, and solve the BVP (\ref{eq:1}), (%
\ref{eq:2}) again using the obtained solution as initial approximation. We
repeat this iteration while at the ``current stage'' $p_{0}$ the equation $%
\lambda _{min}=0$ is satisfied numerically. Then one point at the searched
bifurcation curve is calculated.

At Fig. 2 the typical relationships $\lambda _{min}(\Delta )$ $($here $%
\Delta =2R$ is the junction length) corresponding to the so called ``main
fluxon''\footnote{%
In infinite JJ with one $\delta$-shaped microinhomogeneity at the point $x=0$
``main'' fluxon/antifluxon is represented by the exact solution $\varphi
(x)=4\arctan \exp (\pm x)$ \cite{galpfil}.} (see Fig. 3) in inhomogeneous
junction containing one resistance inhomogeneity placed at the point $x=0$
when $\mu =0.5$ and $\mu =1$ respectively are shown. The points $B_{0}$ and $%
B_{1}$ appear to be bifurcation points of corresponding relationships $%
\lambda _{min}(\Delta )=0$. According to the above mentioned reasonings
these points determine the minimum length of JJ, for which main fluxon is
still stable.

Obviously such kind of algorithm is quite hard. Therefore it is quite
natural to put the question how to calculate directly the bifurcation
curves. A general approach study for the posed problem is proposed at the
articles \cite{boyad88,boyad}. We consider the equations (\ref{eq:1}), (\ref
{eq:2}), (\ref{eq:4}) - (\ref{eq:6}) as a closed nonlinear system with
respect to the functions $\varphi (x),\psi (x)$ and one of parameters $p$
(for example, $h_{B}$ or $\gamma $ \cite{boyad88}), while the other 3
parameters, and $\lambda $ also, are given. Then fixing $\lambda $ to be
small enough, (for example $\lambda =0.01$), every solution of the above
system with a priori prescribed accuracy (the derivative $\dfrac{\partial
\,\lambda }{\partial \,p_{0}}\to \infty $ when $p_{0}$ approaches the its
critical value) belongs to the small vicinity of the searched bifurcation
curve.

The calculation of the critical half-length $R$ of homogeneous or
inhomogeneous JJ, corresponding to a concrete nontrivial distribution of the
magnetic flux, is an important practical problem. The shortcoming in this
connection ensues from the fact that the equations (\ref{eq:1}), (\ref{eq:2}%
), (\ref{eq:4}) - (\ref{eq:6}) are implicit with respect to the quantity $R$%
. At the present work we propose how to overcome the mentioned imperfection.

\section{Method of Solution}

For given values of parameters $\lambda ,$ $\mu ,$ $h_{B}$ and $\gamma $ we
consider the system (\ref{eq:1}), (\ref{eq:2}), (\ref{eq:4})~-(\ref{eq:6})
as a nonlinear eigenvalue problem with respect to the eigenfunctions $%
\varphi (x),\psi (x)$ and to the eigenvalue $R$. As the parameter $R$ does
not occur explicitly , it is convenient to use the Landau transformation $%
\xi =\dfrac{x}{f(R)}$ \cite{vab}. Choosing simply $f(R)\equiv R$ we map the
original interval $[-R,R]$ to the interval $[-1,1]$. Taking into account
that $\dfrac{d}{dx}=\dfrac{1}{R}\dfrac{d}{d\xi }$ we obtain that the above
system renders to the system:

\begin{eqnarray}
-\!\!\!\!\! &&\!\!\!\!\! \bar \varphi _{\xi \xi }+R^{2}[\,\bar j_{D}(\xi
)\sin \bar \varphi +\gamma ]=0\>,  \label{eq:8} \\
&&\!\!\!\!\! \bar \varphi _{\xi }(\pm 1)-R\,h_{B}=0\>,  \label{eq:9} \\
-\!\!\!\!\!&&\!\!\!\!\! \bar \psi _{\xi \xi }+R^{2}[\,\bar j_{D}(\xi )\cos {%
\bar \varphi -\lambda }]\,\bar \psi =0\>,  \label{eq:10} \\
&&\!\!\!\!\! \bar \psi _{\xi }(\pm 1)=0\>,  \label{eq:11} \\
<\!\!\!\!\!&&\!\!\!\!\! \bar \psi ,\bar \psi >-1\equiv R\int_{-1}^{1}\>\bar
\psi^2 (\xi )\,d\xi -1=0\>,  \label{eq:12}
\end{eqnarray}
where $(\bar{.})$ denotes the function of $\xi$. Without fear of confusion
the bars will be omitted henceforth. Let us consider the left-hand side of
the system (\ref{eq:8})-(\ref{eq:12}) as a functional vector $F(\varphi
,\psi ,R)$. Then we have $F(\varphi ,\psi ,R)=0$.

The nonlinear system under consideration will be solved using the continuous
analog of Newton's method \cite{jidk}. Following it, we introduce a
continuous parameter (``time'') $t\in [0,\infty )$ and suppose the
quantities $\varphi ,\,\psi $ and $R$ depend on $t$, i.e., $\varphi (t,\xi
),\,\psi (t,\xi )$ and $R(t).$ The following abstract differential equation
is used

\[
\dfrac{dF}{dt}+F\equiv F_{\varphi }^{\prime }\,\dot{\varphi}+F_{\psi
}^{\prime }\,\dot{\psi}+F_{R}^{\prime }\,\dot{R}+F=0\>. 
\]

Here

\[
F_{\varphi }^{\prime }\,u=\left. \frac{d}{d\epsilon }F(\varphi +\epsilon
\,u)\right| _{\epsilon =0}\>,\,F_{\psi }^{\prime }v=\left. \frac{d}{%
d\epsilon }F(\psi +\epsilon \,v)\right| _{\epsilon =0}\>,\,F_{R}^{\prime
}\,\rho =\left. \frac{d}{d\epsilon }F(R+\epsilon \,\rho )\right| _{\epsilon
=0} 
\]

\noindent are Frechet's derivatives of $F$, and

\begin{equation}
u=\dot{\varphi},\text{ }v=\dot{\psi},\text{ }\rho =\dot{R}  \label{eq:13}
\end{equation}

\noindent are the ``time'' derivatives of the functions $\varphi $, $\psi $
and $R,$ respectively. Simplifying above system we obtain

\begin{eqnarray}
\!\!\!\!\!&-&\!\!\!\!\! u_{\xi \xi }+R^{2}j_{D}(\xi )\cos \varphi
\,u+\{2R\,\,[j_{D}(\xi )\sin \varphi +\gamma ]+R^{2}\,j_{D,\xi }(\xi )\sin
\varphi \}\,\rho  \nonumber \\
\!\!\!\!\!&-&\!\!\!\!\! \varphi _{\xi \xi }+R^{2}[\,j_{D}(\xi )\sin \varphi
+\gamma ]=0\>,  \label{eq:14} \\
&&\!\!\!\!\!u_{\xi }(\pm 1)-\rho \,h_{B}+\varphi _{\xi }(\pm 1)-R\,h_{B}=0\>,
\label{eq:15} \\
\!\!\!\!\!&-&\!\!\!\!\! v_{\xi \xi }+R^{2}\left[ q(\xi )-\lambda \right]
\,v+\left\{ 2R\,\left[ q(\xi )-\lambda \right] \psi +R^2\,\,j_{D,\xi }(\xi
)\,\xi \,\psi \cos \varphi \right\} \,\rho  \nonumber \\
\!\!\!\!\!&-&\!\!\!\!\! R^{2}j_{D}(\xi )\,\psi \sin \varphi \,\,u-\psi _{\xi
\xi }+R^{2}\left[ q(\xi )-\lambda \right] \psi =0\>,  \label{eq:16} \\
&&\!\!\!\!\! v_{\xi }(\pm 1)+\psi _{\xi }(\pm 1)=0,  \label{eq:17} \\
&&\!\!\!\!\! 2R<\psi ,v>+\,\rho <\psi ,\psi >+\,R<\psi ,\psi >-1=0\>,
\label{eq:18}
\end{eqnarray}

This system can be solved using the following decomposition: 
\[
u=u_{1}+\rho \,u_{2}\>,\>\,v=v_{1}+\rho \,v_{2}\>, 
\]

\noindent where $u_{1}(\xi ),u_{2}(\xi ),v_{1}(\xi ),v_{2}(\xi )$ are new
unknown functions of $\xi $. That assumption yields four linear two-point
boundary problems with respect to the new introduced functions:

\begin{eqnarray}  \label{eq:22}
\!\!\!\!\!&-&\!\!\!\!\! u_{1}{}_{\xi \xi }+R^{2}q(\xi )\,u_{1}=\varphi _{\xi
\xi }(\xi )-R^{2}\left[ j_{D}(\xi )\sin \varphi (\xi )+\gamma \right] 
\nonumber \\[-0.2cm]
\\
&&\!\!\!\!\!u_{1}{}_{\xi }(\pm 1)=R\,h_{B}-\varphi _{\xi }(\pm 1)\>, 
\nonumber \\
\nonumber \\
\nonumber \\
\!\!\!\!\!&-&\!\!\!\!\! u_{2}{}_{\xi \xi }+R^{2}q(\xi )\,u_{2} = -2R\,\left[
j_{D}(\xi )\sin \varphi (\xi )+\gamma \right] -R^{2}\,j_{D,\xi }\,(\xi
)\,\xi \sin \varphi (\xi )  \nonumber \\[-0.2cm]
\\
&&\!\!\!\!\! u_{2}{}_{\xi }(\pm 1)=h_{B}\>,  \nonumber \\
\nonumber \\
\nonumber \\
\!\!\!\!\!&-&\!\!\!\!\! v_{1}{}_{\xi \xi }+R^{2}\left[ q(\xi )-\lambda
\right] \,v_{1} =\psi _{\xi \xi }(\xi )-R^{2}\left[ j_{D}(\xi )\,\cos
\varphi (\xi )-\lambda \right] \psi (\xi )+  \nonumber \\
&&\!\!\!\!\!R^{2}\,j_{D,\xi }\,(\xi )\,\psi (\xi )\sin \varphi (\xi
)\,u_{1}(\xi )  \nonumber \\[-0.2cm]
\\
&&\!\!\!\!\! v_{1}{}_{\xi }(\pm 1)=-\psi _{\xi }(\pm 1)\>,  \nonumber \\
\nonumber \\
\nonumber \\
\!\!\!\!\!&-&\!\!\!\!\! v_{2}{}_{\xi \xi }+R^{2}\left[ q(\xi )-\lambda
\right] \,v_{2} = R^{2}j_{D}(\xi )\,\psi (\xi )\sin \varphi (\xi
)\,u_{2}(\xi )-2R\,\left[ q(\xi )-\lambda \right] \psi (\xi )-  \nonumber \\
&&\!\!\!\!\!R^{2}\,j_{D,\xi }(\xi )\,\xi \,\psi (\xi )\cos \varphi (\xi ) 
\nonumber \\[-0.2cm]
\\
&&\!\!\!\!\!v_{2}{}_{\xi }(\pm 1)=0\>.  \nonumber
\end{eqnarray}

At last the norm condition renders to

\begin{equation}
\rho =\dfrac{1-R<\psi ,\psi >-2R<\psi ,v_{1}>}{<\psi ,\psi >+2R<\psi ,v_{2}>}%
.  \label{eq:23}
\end{equation}

For numerical computation the time derivatives $u,v$ and $\rho $ at (\ref
{eq:13}) are discretizated by Euler's method (see details in \cite{jidk}).
For given iteration $k,$ the approximation at the next stage $k+1$ is
obtained as follows

\[
R^{k+1}=R^{k}+\tau ^{k}\rho ^{k}\>,\quad \varphi ^{k+1}=\varphi ^{k}+\tau
^{k}u^{k}\>,\quad \psi ^{k+1}=\psi ^{k}+\tau ^{k}v^{k}. 
\]

Here $\tau ^{k}$ $\in (0,1]$ is a parameter (``time'' step) which can be
chosen satisfying the condition the residual to be minimal (see \cite
{ermakov}).

Let us note that every one of BVP (19)~-(22) can be presented simply in the
form

\begin{eqnarray}
\!\!\!\!\! &-&\!\!\!\!\!y_{\xi \xi }+p(\xi )\,y=r\,(\xi ),\>  \nonumber
\label{eq:24} \\[-0.2cm]
&& \\
&&\!\!\!\!\!y_{\xi }(\pm 1)=y_{\pm ,}  \nonumber
\end{eqnarray}

\noindent where $p(\xi )$, $r\,(\xi )$ are given functions, and $y_{\pm }$
are given constants. Further we define an uniform set at the interval $%
[-1,1] $ namely $\xi _{j}=-1+jh\>,\,\>j=0,\ldots ,N$, where $N$ is number of
knots, $h=\dfrac{2}{N}$ is the step of set. Let $S(\xi )$ is a cubic spline
interpolating the function $y(\xi )$ over the $\xi $-mesh. We assume that $%
M(\xi )=:S_{\xi \xi }(\xi )$. Then taking into account the continuity
condition concerning the first moments of the spline $S(\xi )$ and BVP (\ref
{eq:24}), we obtain a three-diagonal algebraic system (see \cite{zavyal}).

The iteration process starts from the initial conditions $\varphi
_{j}^{0}=\varphi _{j}^{k},\psi _{j}^{0}=\psi _{j}^{k},$ $j=1,\ldots N$ and $%
R^{0}=R^{k}$, where $k$ denotes the number of the iteration. We solve
consequently the two-point BVP (19)~-(22). Thus the values for the grid
functions $u_{1j}^{k}\>,u_{2j}^{k}\>,v_{1j}^{k}\>,v_{2j}^{k}$ are obtained.
Then using (\ref{eq:23}) we calculate the increment $\rho $. Hereupon we
calculate the increments $u$ and $v$ and obtain the predictions for the
junction length $R^{k+1}$ and the grid functions $\varphi _{j}^{k+1}\>,\psi
_{j}^{k+1}$ at the new stage $k+1$. The criterion for terminating the
iteration is

\[
\max_{j}\,(\delta \varphi ,\delta \psi ,\delta R)\le \varepsilon \>, 
\]

\noindent where $\delta \varphi ,$ $\delta \psi $ and $\delta R$ are the
corresponding residuals. We use the norm estimation $\varepsilon \sim
10^{-8}\div 10^{-12}$.

\section{Results and Discussion}

The numerical correctness of the used scheme is verified through appropriate
numerical experiments. We used different meshes with sizes $N=256,512$ and $%
1024$. The relative differences for the magnetic flux $\varphi (x)$, first
eigenfunction $\psi (x)$ do not exceed $0.004\%$ and $0.03\%$, respectively.
We estimated the order of approximation of the obtained solution using the
Runge method. The calculations carried out on meshes with spacings $h=\dfrac{%
1}{128},\dfrac{1}{256},\dfrac{1}{512}$ are presented on the table. It is
easy to show the approximate relationship 
\[
\dfrac{z_{h}-z_{\frac{h}{2}}}{z_{\frac{h}{2}}-z_{\frac{h}{4}}}\approx 2^{2} 
\]
holds at the knots $0;\frac{N}{2};N$. Here $z=\left\{ \varphi ,\psi
,R\right\}.$ On the ground of that comparison we conclude the second-order
approximation for the functions $\varphi (x)$, $\psi (x)$ and the junction
half-length $R$ is satisfied.

\bigskip

\begin{center}
{\small 
\begin{tabular}{||c|c|c|c|c|c|c|c||}
\hline
$N$ & $h$ & $\varphi _{0}\equiv \varphi (-R)$ & $\varphi _{\frac{N}{2}%
}\equiv \varphi (0)$ & $\varphi _{N}\equiv \varphi (R)$ & $\psi _{0}\equiv
\psi _{N}\equiv \psi (\pm R)$ & $\psi _{\frac{N}{2}}\equiv \psi (0)$ & $%
R_{min}$ \\ \hline
$256$ & $\frac{1}{128}$ & $1.1770473$ & $3.1415927$ & $5.1061380$ & $%
0.4203551$ & $0.5178962$ & $2.1145618$ \\ \hline
$512$ & $\frac{1}{256}$ & $1.1770277$ & $3.1415927$ & $5.1061576$ & $%
0.4202415$ & $0.5177725$ & $2.1145703$ \\ \hline
$1024$ & $\frac{1}{512}$ & $1.1770244$ & $3.1415927$ & $5.1061609$ & $%
0.4202198$ & $0.5177485$ & $2.1145694$ \\ \hline
\end{tabular}
}
\end{center}

\bigskip

All results, stated below, are related to the solutions of kind ``main
fluxon'' (see Fig. 3).

The upper curve on the Fig.4 presents the initial distribution of the
magnetic field $\varphi _{x}(x)$ alongside the junction. This distribution
is chosen to be a solution of BVP (\ref{eq:1}), (\ref{eq:2}) for $R=5$, $%
h_{B}=0$ and $\gamma =0$. At the same figure the lower curve presents the
calculated distribution of the magnetic field corresponding to the minimal
eigenvalue $\lambda _{min}=0.01$. Hence this is the searched bifurcation
distribution and in the framework of this model the value of the spectral
parameter $R_{min}\approx 2.11$ is the minimal half-length of the junction
providing a stable main fluxon. In this sense, junctions, whose length lies
below the critical $R_{min}$, it is necessary to be considered as ``short''
for distributions of the kind ``main fluxon''. As it was shown \cite{boyad85}%
, for such short length, there is a unique stable distribution of the
magnetic flux in the junction - Meissner's solution\footnote{{\normalsize %
This is the trivial solution $\varphi (x)=0$ (stable) and $\varphi (x)=\pi $
(unstable) for $h_{B}=0$ and $\gamma =0.$}}.

On Fig.5 the distributions of magnetic field $\varphi _{x}(x)$ alongside the
junction in absence of external current ($\gamma =0$) depending on its
boundary values $h_{B}=0$ and $h_{B}=1$ respectively, are represented. It is
seen that two solutions could be approximated by means of two-degree
polynomials. Furthermore they seem to be geometrically similar. Namely the
curve $\varphi _{x}(x)$ corresponding to magnetic field $h_{B}>0$ may be
considered as obtained from the other curve doing a homothety alongside axis 
$x$ with respect to the pole $x=0$ and a translation alongside the axis $%
\varphi _{x}(x)$.

On Fig.6 the obtained relationship $\lambda _{min}(h_{B},\gamma )=0$ is
drawn, depending on the length of the trapezium base $\mu $. It is well
noticeable the stabilizing influence of the boundary magnetic field upon the
critical length of the junction, i.e., increasing the magnitude of above
mentioned quantity one decreases the critical length of junction $2R_{min}$.
Thus the calculated critical half-length of the Josephson junction
corresponding to the so called main soliton ($h_{B}=0,$ $\gamma =0$) is $%
R_{min}\approx 2.11$ while the same quantity when $h_{B}=1$ and $\gamma =0$
is $R_{min}\approx 1.35$.

\section{Concluding Remarks}

A direct iterative method for obtaining the minimal half-length $R_{min},$
corresponding to fixed distribution of the magnetic flux in a long Josephson
junction is developed. The mere existence of the minimal length is a
warranty enough for us to name ``long'', every JJ in which there exists at
least one nontrivial stable distribution of the magnetic flux for given
values of other parameters.

An appropriate linearization based on the continuous analog of Newton's
method renders the original nonlinear spectral problem to four two-point
linear BVP. A spline-difference scheme in second order of approximation for
solving of these BVP is used.

This method may be applied for solving more general nonlinear problems.

\section{\protect\bigskip Acknowledgments}

We thank Prof. I.V. Puzynin (JINR, Dubna, Russia) for useful remarks.

This research was supported by the Bulgarian Ministry of Education, Science
and Technologies under Grants MM-425/94 and MM-602/96.

\bigskip

%\bibliography{todor}

\begin{thebibliography}{99}
\bibitem{alexboyad}  N.~V. Alexeeva and T.~L. Boyadjiev. \newblock Periodic
bound states in {J}osephson lattices of resistive inhomogeneities. 
\newblock 
\emph{Bulg. J. of Physics}, (No 1,2), 1997.

\bibitem{barone}  A.~Barone and G.~Paterno. \newblock \emph{Physics and
Apllications of the Jisephson Effect}. \newblock John Wiley \& Sons, Inc.,
1920.

\bibitem{boyad85}  T.~L. Boyadjiev, Yu.~S. Gal'pern, I.~V. Puzynin, and
A.~T. Filippov. \newblock Bifurcations of bound states of fluxons in
inhomogeneous perturbed by bias current and external magnetic field. %
\newblock \emph{Comm. of JINR}, P11-85-807, Dubna, 1985.

\bibitem{boyad86}  T.~L. Boyadjiev, Yu.~S. Gal'pern, I.~V. Puzynin, and
A.~T. Filippov. \newblock Bound states of fluxons in inhomogeneous {J}%
osephson junction {J}osephson junction of finite length. \newblock \emph{%
Comm. of JINR}, P17-86-506, Dubna, 1986.

\bibitem{ermakov}  V.~V. Ermakov and N.~N. Kalitkin. \newblock Optimal step
and regularization of {N}ewton's method. \newblock \emph{JVMiMF}, 21(2),
1981. \newblock in Russian.

\bibitem{vystavkin}  A.~N.~Vystavkin, Yu. F. Drachevski, V. P. Kosheletz and
I. L. Serpuchenko. \newblock Detection of static bound states in the
distributed Josephson junctions with inhomogeneities\newblock \emph{Fizika
nizkih temperatur}, 14(No 6):646--649, 1988. \newblock in Russian.

\bibitem{jidk}  E.~P.~Jidkov, G. I. Makarenko and I. V. Puzynin. \newblock %
Continuous analog of {N}ewton's method in nonlinear physical problems. %
\newblock In \emph{Physics of elementary particles abd nuclei}, number v.4,
No 1, pages 127--166, Dubna, 1973. \newblock in Russian.

\bibitem{boyad88}  T.~L.~Boyadjiev, D. V. Pavlov and I. V. Puzynin. %
\newblock Newton's algorithm for calculation of critical parameters in
one-dimensional inhomogeneous Josephson junctions. \newblock \emph{Comm. of
JINR}, P11-88-409, Dubna, 1988. \newblock in Russian.

\bibitem{galpfil}  Yu.~S. Gal'pern and A.~T. Filippov. \newblock Bound
states of solitons in inhomogeneous Josephson junctions. \newblock \emph{JETF%
}, 86(4):1527--1543, 1984. \newblock in Russian.

\bibitem{vab}  P.~N. Vabishchevich. \newblock \emph{Numerical methods for
solving free-boundary problems}, chapter~3, pages 59--60. \newblock %
Publishing House of the Moscow State University, 1987. \newblock in Russian.

\bibitem{zavyal}  Yu. S. Zavyalov, B.~I. Kvasov and V.~L. Miroshnichenko. %
\newblock \emph{Methods of spline-functions.} \newblock Nauka., Moscow,
1980. \newblock in Russian.
\end{thebibliography}

\section{\protect\bigskip FIGURE CAPTIONS}

Figure 1: $a)$ Geometrical sketch of an inhomogeneous JJ; $b)$ Geometrical
model of the amplitude $j_{D}(x)$ of Josephson current.

\bigskip Figure 2: The minimal eigenvalue $\lambda _{min}$ as function of
the junction length $\Delta =2R$.

\bigskip Figure 3: The magnetic flux $\varphi (x)$, magnetic field $\varphi
_{x}(x)$ and first eigenfunction $\psi (x)$ as functions of the junction
length $\Delta =2R$.

\bigskip Figure 4: The magnetic field $\varphi _{x}(x)$ alongside the
junction as function of its length $2R$: $"\nabla "$ - initial distribution;
''$\circ "$ - final (bifurcation) distribution, corresponding to the minimal
length.

\bigskip Figure 5: The magnetic field $\varphi _{x}(x)$ alongside the
junction as function of its length $2R$ for length of the trapezium base $%
\mu =1$, bias current $\gamma =0$ and different values of the boundary
magnetic field $h_{B}$: $"\circ "$ - $h_{B}=0\>$; $"\diamond "$ - $h_{B}=1\>$%
.

\bigskip Figure 6: The minimal (bifurcation) eigenvalue $\lambda _{min}$ as
a function of the boundary magnetic field $h_{B}$ for different lengths of
the trapezium base $\mu $: ''$\nabla "$ - $\mu =0.5\>$; $"\triangle "$ - $%
\mu =1\>$.

\end{document}